\documentclass[doublecol]{epl2} 
\usepackage{mathptmx,amssymb}
\usepackage{amsmath}

\title{Bifurcations of a large scale circulation in a quasi-bidimensional turbulent flow}
\shorttitle{Bifurcations of a large scale circulation on a turbulent background} 

\author{G. Michel \and J. Herault   \and F. P\'etr\'elis \and S. Fauve} 
\shortauthor{G. Michel \etal}

\institute{                    
  \inst{1} Laboratoire de Physique Statistique, Ecole Normale Sup\'erieure, CNRS, Universit\'e P. et M. Curie, Universit\'e Paris Diderot, Paris, France\\
}

\abstract{We report the experimental study of the bifurcations  of a large-scale circulation that is formed over a turbulent flow generated by a spatially periodic forcing.  After shortly describing how the flow becomes turbulent through a sequence of symmetry breaking bifurcations, we focus our study on the transitions that occur within the turbulent regime. They are related to changes in the shape of the probability density function (PDF) of the amplitude of the large scale flow. We discuss the nature of these bifurcations and how to model the shape of the PDF.}

\begin{document}

\maketitle

{\bf Introduction\\}


Experiments on nearly two-dimensional flows generated in a thin layer of fluid by a spatially periodic forcing have been first performed to study the generation of large scale flows \cite{bondarenko:1979, tabeling:1987} and the properties of two-dimensional turbulence \cite{Sommeria,paret:1997,ref1,ref2}. These flows have been modeled using the two-dimensional Navier-Stokes equation with an additional term describing fluid friction on the bottom boundary of the fluid layer. Thus, besides the Reynolds number, $Re$, a second dimensionless parameter, $Rh$, describes the ratio of the inertial force to fluid friction. The stability of the linear response of the flow to different forcing geometries has been investigated analytically \cite{thess:1992} and the sequence of bifurcations leading to a chaotic behavior in space and time has been studied using numerical simulations \cite{braun:1997, pankaj}.
It has been found that the first instabilities observed as the control parameters ($Re$ and $Rh$) are increased, break the planar symmetries of the forcing and give rise to time periodic and quasi periodic behaviors. As the control parameters are increased further, chaos is observed and the system explores in phase-space one among different attractors  (images under planar symmetries). Then, these attractors merge through a crisis and this gives rise to a symmetric attractor. Thus planar symmetries are statistically restored. 

It is indeed a common belief that in strong enough turbulent regimes, the system explores the whole available phase space because turbulent fluctuations trigger transitions between symmetric attractors. However, several examples of transitions between different turbulent regimes have been reported, for instance the drag crisis that corresponds to a  modification of the geometry of the turbulent wake behind a sphere or a cylinder \cite{Tritton:1977}, the generation of a large scale flow in turbulent convection \cite{Krishnamurti:1981}, broken symmetries in turbulent von Karman swirling flows \cite{Labbe:1996,Ravelet:2004}.  In  the case of two-dimensional confined flows, it has been predicted \cite{kraichnan:1967,rob}  and observed \cite{Sommeria,paret:1997} that the inverse cascade of energy either leads to an homogeneous turbulent flow displaying a wide range of wave numbers $k$ with a $k^{-5/3}$ scaling law, or to a condensate regime that results from an accumulation of kinetic energy in the lowest mode of the fluid layer.  The transition between the two regimes depends on the importance of large scale friction, i.e. on the value of $Rh$. In the presence of friction, the inverse cascade stops at a length scale $l_I$. When $Rh$ increases, $l_I$ reaches the size of the fluid layer $L$ and energy accumulates at the lowest wave number, thus leading to a dominant large scale circulation. 

Here we study the evolution of a large scale mode in a configuration with a small scale separation. After shortly describing the sequence of bifurcations that leads to a turbulent regime when $Rh$ is increased for $Re$ large, we report in this letter an experimental study of the transitions that occur within the turbulent regime when $Rh$ is increased further. We first observe that the probability density function (PDF) of the large scale velocity changes from Gaussian to bimodal when $Rh$ is increased. Above a critical value $Rh_6$, we show that the PDF can be fitted by the superposition of two symmetric Gaussians with a separation between their mean values increasing like $\sqrt{Rh-Rh_6}$ and a nearly constant standard deviation. The bimodality becomes more and more pronounced as $Rh$ is increased and is related to random reversals of the large scale circulation. The average waiting time between successive reversals becomes longer and longer and a condensed regime with no reversal is finally observed. Thus, the regime with random reversals of the large scale circulation is located in parameter space between the condensed state and the turbulent regime with Gaussian large scale velocity. \\

{\bf Experimental set-up and techniques\\}


A thin layer of  liquid metal (Galinstan) of thickness $h=2\,$cm, is contained in a square cell of length $L=\, 12$ cm and is subject to a uniform vertical magnetic field up to $B_0 \simeq 0.1$T. A DC current $I$ (0-200A) is injected at  the bottom of the cell through an  array of $2\times 4$ electrodes (see \cite{jojoepl2015} for a more detailed description of the experiment). The Lorentz force associated to
the current and the magnetic field drives a cellular flows, described
in the next section.

In addition to $h$, $L$, $B_0$ and $I$, the relevant physical parameters are the fluid density $\rho=6.44\,$ kg.m$^{-3}$, its viscosity $\nu=3.72\, 10^{-7}\,$ m$^2$.s$^{-1}$, its electrical conductivity $\sigma=3.46\, 10^{6}\,$ S.m$^{-1}$ and the magnetic permeability of vacuum $\mu_0$. 
Four independant dimensionless numbers can therefore be constructed based on these parameters. However two of them can be ignored since the magnetic field is strong and the flow speed is small. 
More precisely, in the limit of large Hartmann number $\hbox{Ha}= h B_0  [\sigma/(\rho \nu)]^{1/2} \sim 10^2$ and small magnetic Reynolds number  $Rm=\sigma \mu U_c L \sim 10^{-2}$ (with $U_c$ the characteristic speed of the flow), the velocity field is nearly two dimensional, and its vertical average satisfies the two dimensional Navier-Stokes equation with an additional linear damping term $-\textbf v/ \tau_H$ with 
$\tau_H=h^2/(\nu \hbox{Ha})$ \cite{ns2D}.

This quasi-2D flow depends on  two dimensionless numbers, {\it e.g.} the usual  Reynolds number $Re=U_c L/\nu$ and $Rh=U_c \tau_H/L$ which is the ratio of inertia to linear friction. For large $Re$ and $Rh$, the characteristic velocity $U_c$ is set by a balance between inertia and  the Lorentz force and reads $U_c=\sqrt{I B_0/(\rho h)}$. The ratio $Re/Rh$, independent of the injected current, is equal to  $\hbox{Ha} (L/h)^2 \sim 10^4$. By changing $I$ we vary $Rh$ between $1$ and $50$.  Since viscous dissipation becomes efficient at scales smaller than $l = L \sqrt{Rh /R e} \sim 10^{-3}\,$ m,  dissipation at large scale is mainly due to the friction term. It follows from these order of magnitude estimates that $Rh$ is the relevant control parameter for  the large scales dynamics, which is well verified experimentally~\cite{herault}.

In the following we focus on the behavior of the large scale velocity component measured by  the potential difference between a pair of electrodes in the external magnetic field~\cite{cramer2006}.  One of the electrodes is located in the middle of the cell and the other one close to the lateral wall, $5$ mm away from it  { (see figures in \cite{jojoepl2015}).  The flow  induces an electromotive force $\Delta V=\int_{L/2} (\textbf u \times \textbf B_0) . {\bf d l}\simeq \phi_L B_0/h$ where $L/2$ is the distance between the two electrodes and $\phi_L$ is  the flow rate between the center and the wall.} From now on, we consider the spatially averaged velocity  normalized by $U_c$, {\it i.e.} $V=2 \phi_L/(h\, L\, U_c)$,   which is therefore the large scale velocity coarse-grained on size $L/2$. \\


\begin{figure*}[htb!]
\begin{center}
\includegraphics[width=170mm]{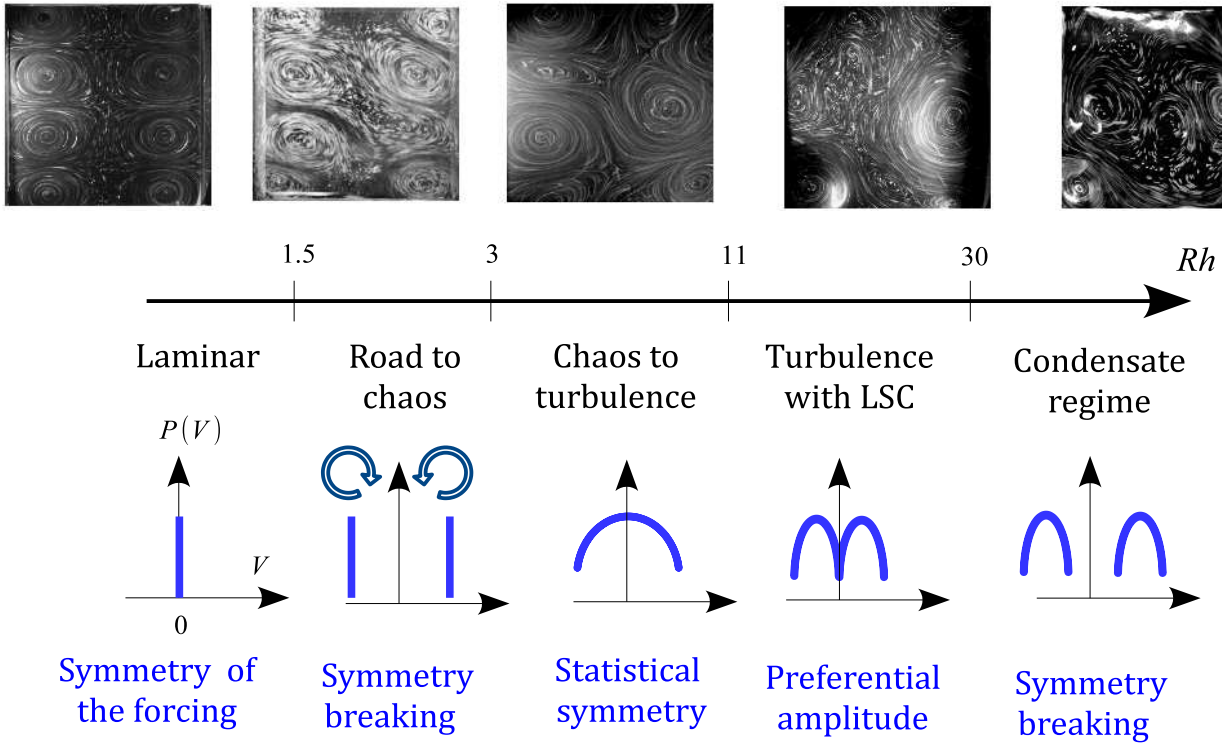} 
\caption{\label{fig1} Top: Pictures of the flow as a function of $Rh$. From left to rigth: laminar flow, first bifurcation, chaotic flow, turbulent flow with moderate large scale flow, turbulent flow with strong large scale flow (condensate). Bottom: Sketch of the PDF of large scale circulation. Symmetry breaking indicates that depending on the initial conditions one of the two states ({\it i.e. } one of the two peaks of the PDF) will be observed. }
\end{center}
\end{figure*}

\begin{figure} 
\begin{center}
\includegraphics[width=70mm,height=45mm]{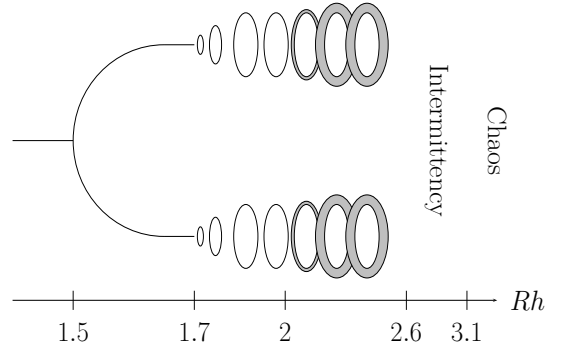} 
\caption{\label{fig2} Bifurcation diagram as a function of $Rh$. The base state is unstable toward a steady state through a direct pitchfork bifurcation. It is followed by two direct Hopf bifurcations. The generated torus (in phase-space) is unstable and an intermittent regime appears. At larger $Rh$, the time series do not display anymore the intermittent phase.}
\end{center}
\end{figure}

{\bf First bifurcations, appearance of a chaotic regime\\}

A summary of the evolution of the velocity field is presented in fig. 1 together with pictures of the flow. 
In the vicinity of each electrode (diameter $d=8\, \hbox{mm}$ flush to the bottom of  the fluid layer),  
the current density $\textbf j$ is radial so that the associated   Lorentz force density $\textbf f_L=\textbf j \times \textbf B_0$  creates a  local  torque.  For $Rh\le Rh_1=1.55$, this forcing drives a laminar flow made of an  array of 8 counter-rotating vortices. The large scale velocity vanishes as expected from the symmetries of the forcing. Increasing  $Rh$ makes this base flow  unstable. 
\begin{itemize}
\item A first bifurcation takes place at  $Rh_1$. The
streamlines of one of the two pairs of diagonal  vortices merge, leading
to the appearance of a large scale circulation through a
direct steady pitchfork bifurcation. Consequently the amplitude of the large scale flow increases proportionally to  $(Rh-Rh_{1})^{1/2}$ (see fig. \ref{figmesurebif}).

\item For slightly larger $Rh$, at  $Rh_{2}=1.7$, a secondary bifurcation takes place. The flow becomes time-dependent. This is a direct Hopf bifurcation with period at onset $T_2 = 4.2\, s$. The amplitude of the corresponding  Fourier mode  increases  proportionally to $(Rh-Rh_{2})^{1/2}$. 

\item A second Hopf bifurcation takes place at $Rh_{3}=2$. Its period $T_3$ ($1.1\, s$ at onset) cannot be expressed as  a simple rational number times $T_2$ (the two periods vary continuously and with opposite monotony with $Rh$). There is thus no frequency-locking between these two oscillations. In phase-space, a torus is generated, as sketched in fig. 2. Here again, the amplitude of the corresponding  Fourier component  increases  proportionally to $(Rh-Rh_{3})^{1/2}$.

\begin{figure}
\begin{center}
\includegraphics[width=70mm,height=45mm]{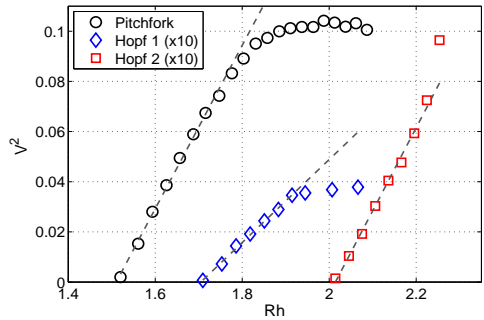} 
\caption{\label{figmesurebif} Square of the amplitude of the unstable mode for the first three bifurcations as a function of $Rh$. The squared amplitudes of both Hopf bifurcations have been multiplied by $10$. }
\end{center}
\end{figure}

\item The following bifurcation, at $Rh_{4}\simeq 2.6$, corresponds to a destabilization of this torus. In phase space, the system spends long durations close to this torus, but is unstable: a component at a non commensurate frequency increases and the time series is chaotic before the system is re-injected close to the torus. This is similar to the second Pomeau-Manneville scenario of intermittent transition to chaos describing
the destabilization of a limit cycle \cite{PM}. 

Slightly further above this transition, the attractor in phase-space explores both signs of the amplitude of the large scale circulation. The symmetry of the forcing is statistically recovered. Increasing $Rh$ above $Rh_5 \simeq 3.1$, the time series become more chaotic and the laminar phases disappear. \\

\begin{figure}
\begin{center}
\includegraphics[width=80mm]{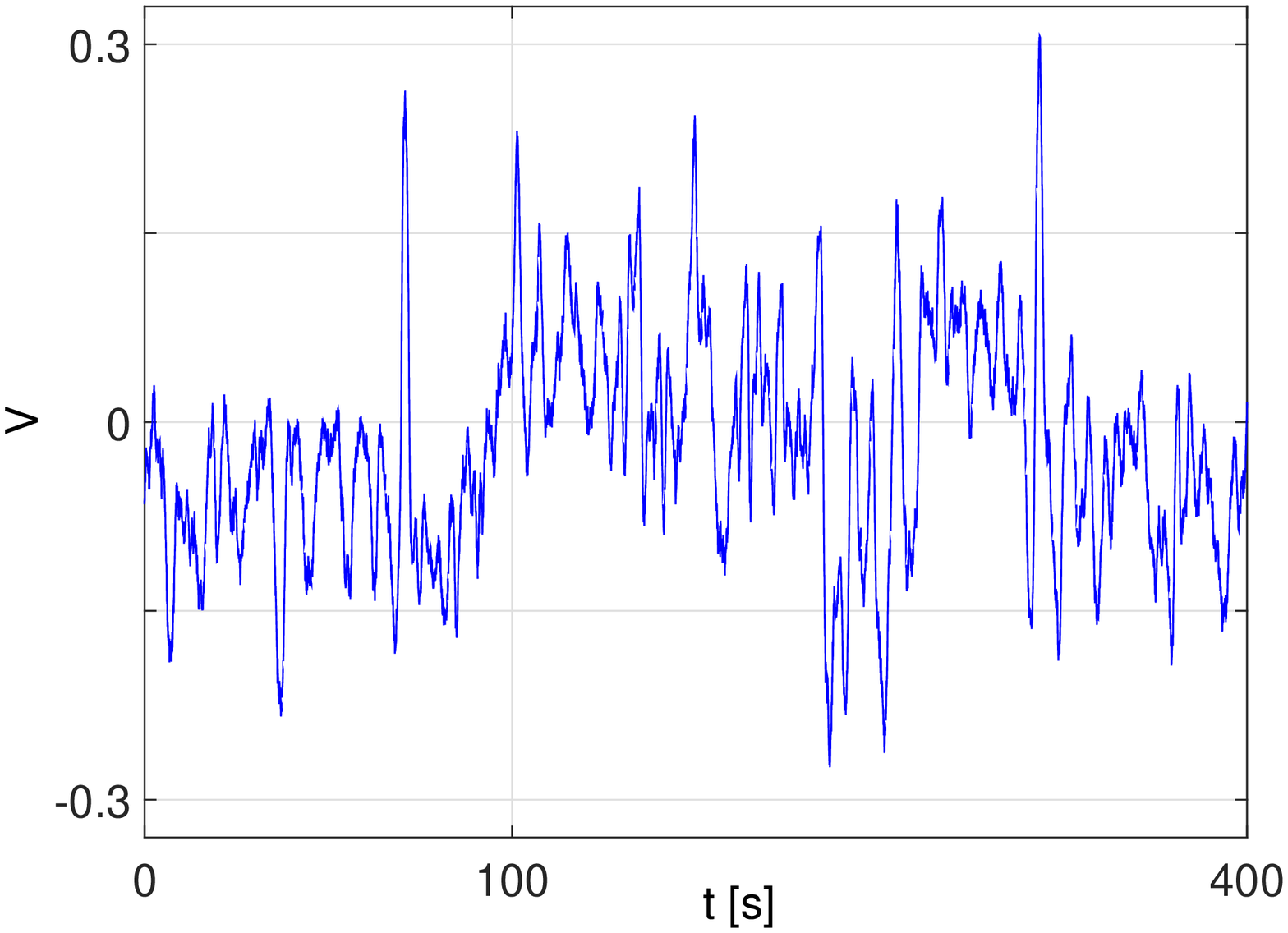}  
\includegraphics[width=80mm]{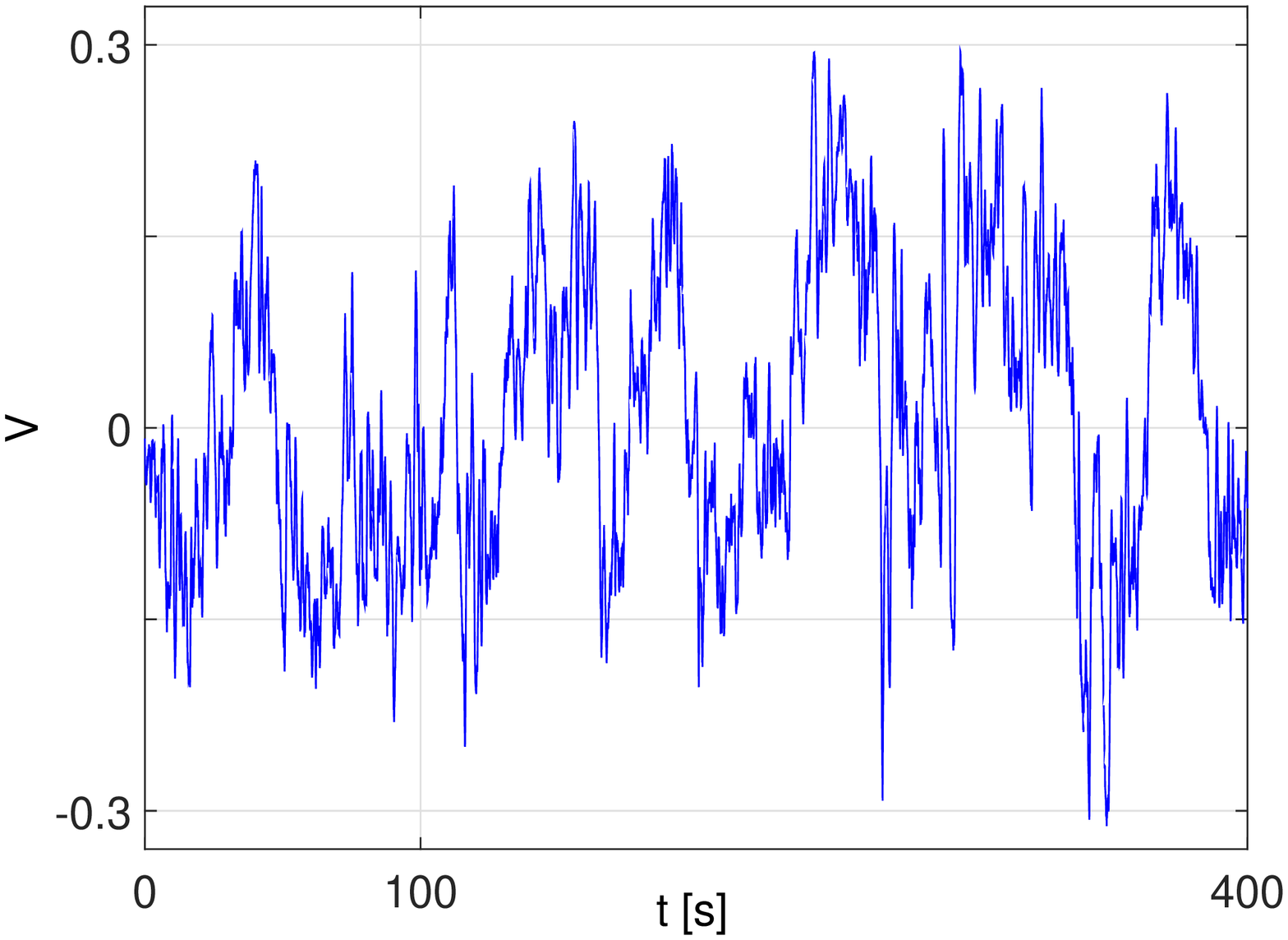}  
\caption{\label{figser} Time series of the large scale circulation at  (top) $Rh=10$ and (bottom) $Rh=20$.}
\end{center}
\end{figure}

\begin{figure}
\begin{center}
\includegraphics[width=70mm,height=45mm]{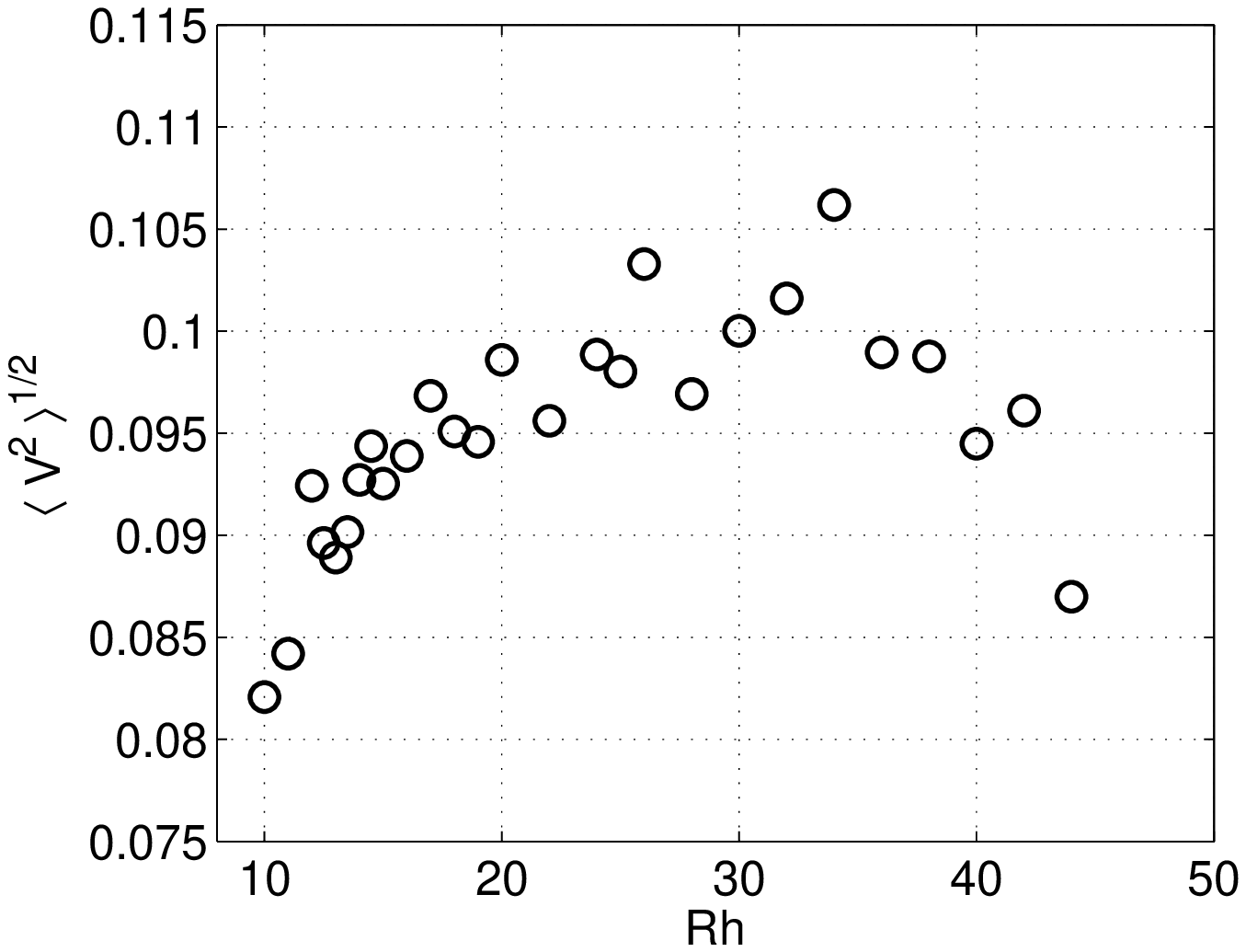} 
\caption{\label{figstdUl} Amplitude of the large scale velocity $\langle V^2\rangle^{1/2}$ as a function of $Rh$.}
\includegraphics[width=70mm,height=45mm]{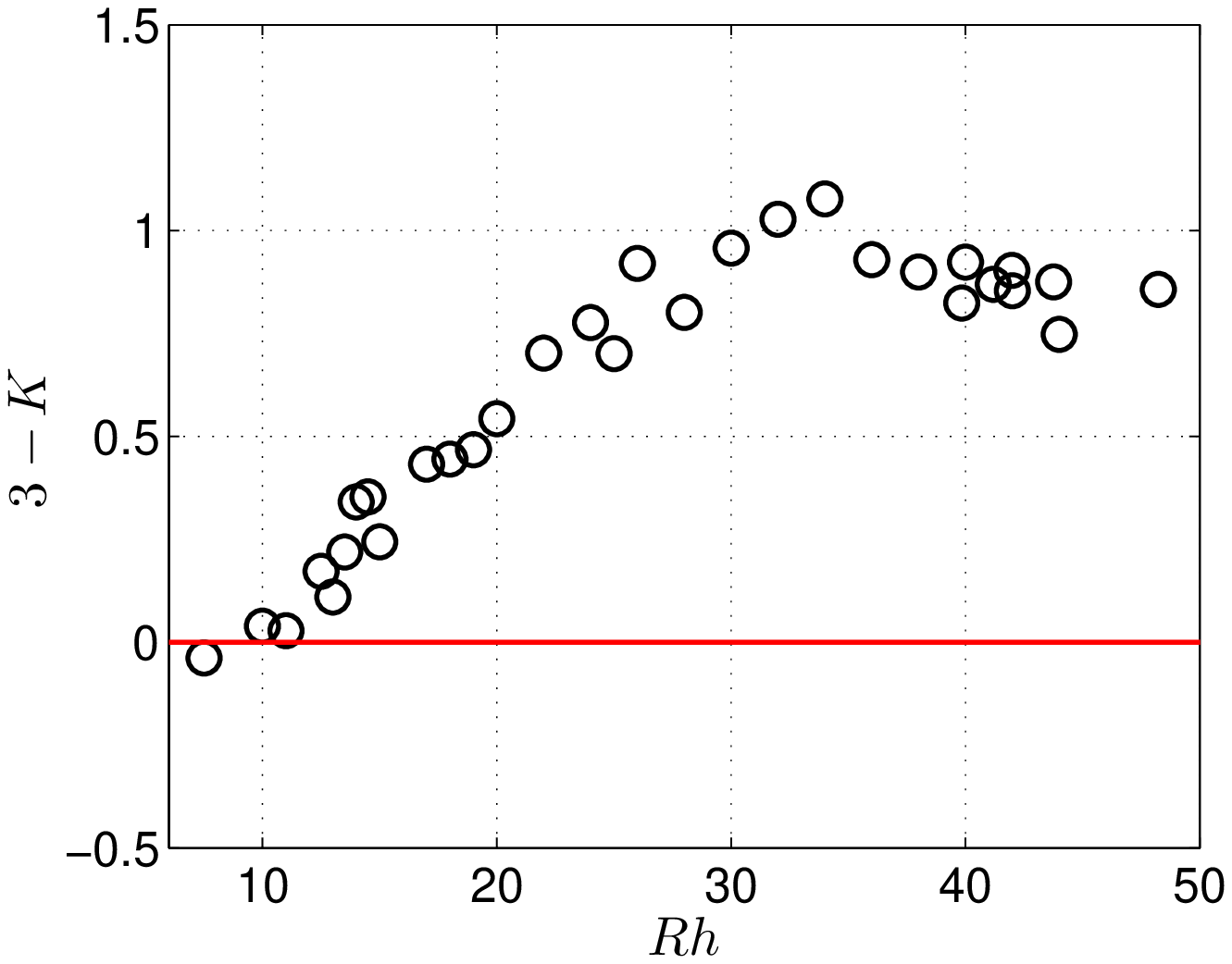} 
\caption{\label{figkurto} Kurtosis of the large scale velocity field: $\langle V^4 \rangle/\langle V^2\rangle^2$. }
\end{center}
\end{figure}

\end{itemize}

For all these transitions, bifurcation theory provides scaling laws once the broken symmetry is known. Many hydrodynamics system going from a laminar to a chaotic state can be characterized in such a way,  well-known examples being Rayleigh-Benard convection or Taylor-Couette flow. \\

{\bf Next bifurcations, appearance and disappearance of reversals\\ }

For $Rh$ slightly below $Rh_6 \simeq 11$, the amplitude of the large scale circulation fluctuates around zero. { We stress that the flow is turbulent in the following sense: even though scale separation between the forcing scale and the container size is not large, the temporal spectrum of the velocity field displays a continuous part over more than three decades in frequency. In addition, energy can be measured  up to large wavevectors due to the direct cascade of enstrophy. Note also that visualization of the large scale vorticity \cite{herault} displays interacting vortices of different sizes.”}

The  next transition is related to the appearance of coherent states where the flow maintains its direction for long duration. Such events are visible in fig \ref{figser} (bottom), while at lower $Rh$ (top), the amplitude is most of the time close to zero and displays short bursting events.  

We now want to quantify the appearance of this non zero state, that would correspond to the amplitude of the unstable mode in usual bifurcations. This is not straightforward as the basic state is already turbulent. 
We thus rely on statistical properties of the signal. 
The variance ($\langle V^2\rangle$) and the Kurtosis ($\langle V^4\rangle/\langle V^2\rangle^2$) are displayed in Fig \ref{figstdUl} and \ref{figkurto}. We note that at $Rh_{6} \simeq 11$, a transition occurs. The distribution is Gaussian ($K=3$) for $Rh\le Rh_{6}$ and becomes non Gaussian above $Rh_{6}$.

To obtain a finer description of the bifurcation, we have to consider the  PDF of $V$. 
We point out that this can be technically difficult: in such experiments, any imperfection  would bias the system even slightly toward one direction of rotation (i.e. one sign of $V$).  
Several analyses presented hereafter, in particular involving fits, are highly sensitive to any asymmetry of the system. For such analysis,  we restrict to experiments for which the system is well equilibrated and the distributions are symmetrical.

As displayed in fig. 7, the shape of the PDF changes with $Rh$. It is close to a 
Gaussian at $Rh=12$, it is  flatter at its center  at $Rh=14$ and, at $Rh=20$,
it is bimodal with a local minimum at $V=0$.  The evolution of the PDF traces back
to the modification of the time series of $V$: the appearance of non zero
temporary attractive states is responsible for the bimodal structure of the
PDF.

The evolution of the shape of the PDF can be captured using the following
model. We consider that it is the sum of two Gaussian of width $\sigma$ and
centered at $\pm dX$. We write
\begin{eqnarray}
P[V] = & (2 \sqrt{2 \pi} \sigma)^{-1} \large( exp(-(V-dX)^2/(2 \sigma^2))\nonumber \\
&+exp(-(V+dX)^2/(2 \sigma^2))\large) \,. \label{defP1}
\end{eqnarray}

For $Rh$ below $Rh_6$ a single Gaussian provides a good fit to the PDF,  and this
corresponds to $dX=0$.
For larger $Rh$, we extract $dX$ and $\sigma$ from best fits of the whole PDF. They are displayed in fig. \ref{figumsigma}. 
Note in particular that both the center and the tails of the PDF are well
fitted by Eq. (\ref{defP1}).
The standard deviation $\sigma$ of each PDF  remains nearly constant, whereas its 
center, $dX$, increases with $Rh$. As displayed in the insert of
fig. \ref{figumsigma}, this behavior is compatible with a power law $(Rh-Rh_6)^{1/2}$. 
As often when trying to extract critical exponents, we note
that due to the error bars, our measurements do not exclude values of the
exponent close to but different from $1/2$.


The PDF becomes bimodal at a larger value $Rh_7\simeq 17$.  This corresponds to
$dX=\sigma$, the Rayleigh criterion  for separating
two lines in an optical spectrum. This secondary bifurcation of the PDF, associated to the appearance of bimodality is similar to the one of the free-energy in the context of second-order phase transition in the model of Landau.  Led by this analogy,  in the vicinity of the transition (here close to $Rh_7$),  we model the PDF as
\begin{equation}
P[V] \propto exp{(a V^2+b V^4)}\,.
\label{eqlandau}
\end{equation}
We emphasize that this model is restricted to small values of $V$ (it is not expected to model the tails of the PDF). Landau's assumption is that $a$ varies linearly in the control parameter and changes sign at the transition while $b$ remains roughly constant. 
 
\begin{figure}
\begin{center}
\includegraphics[width=75mm]{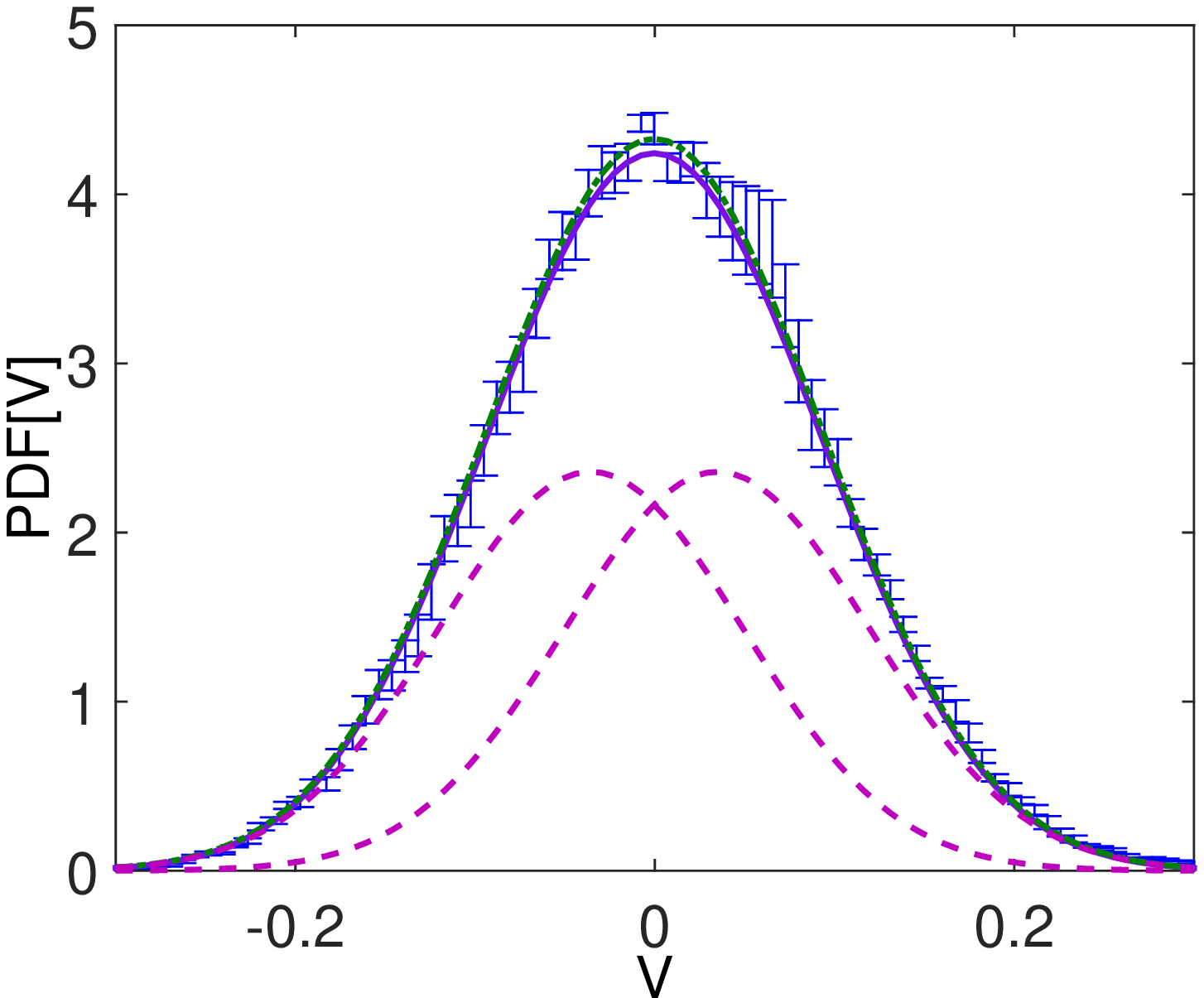}   
\includegraphics[width=75mm]{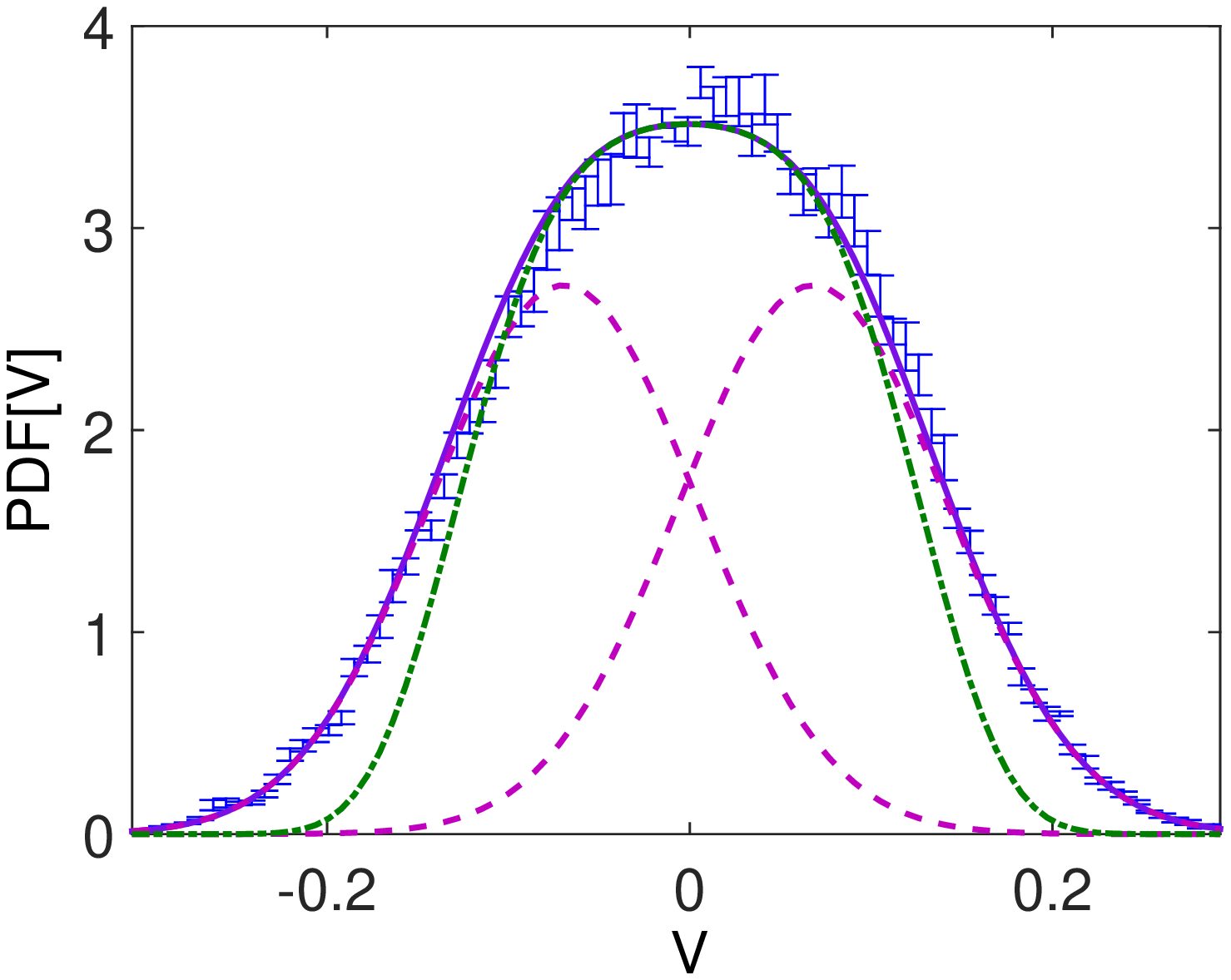}  
\includegraphics[width=75mm]{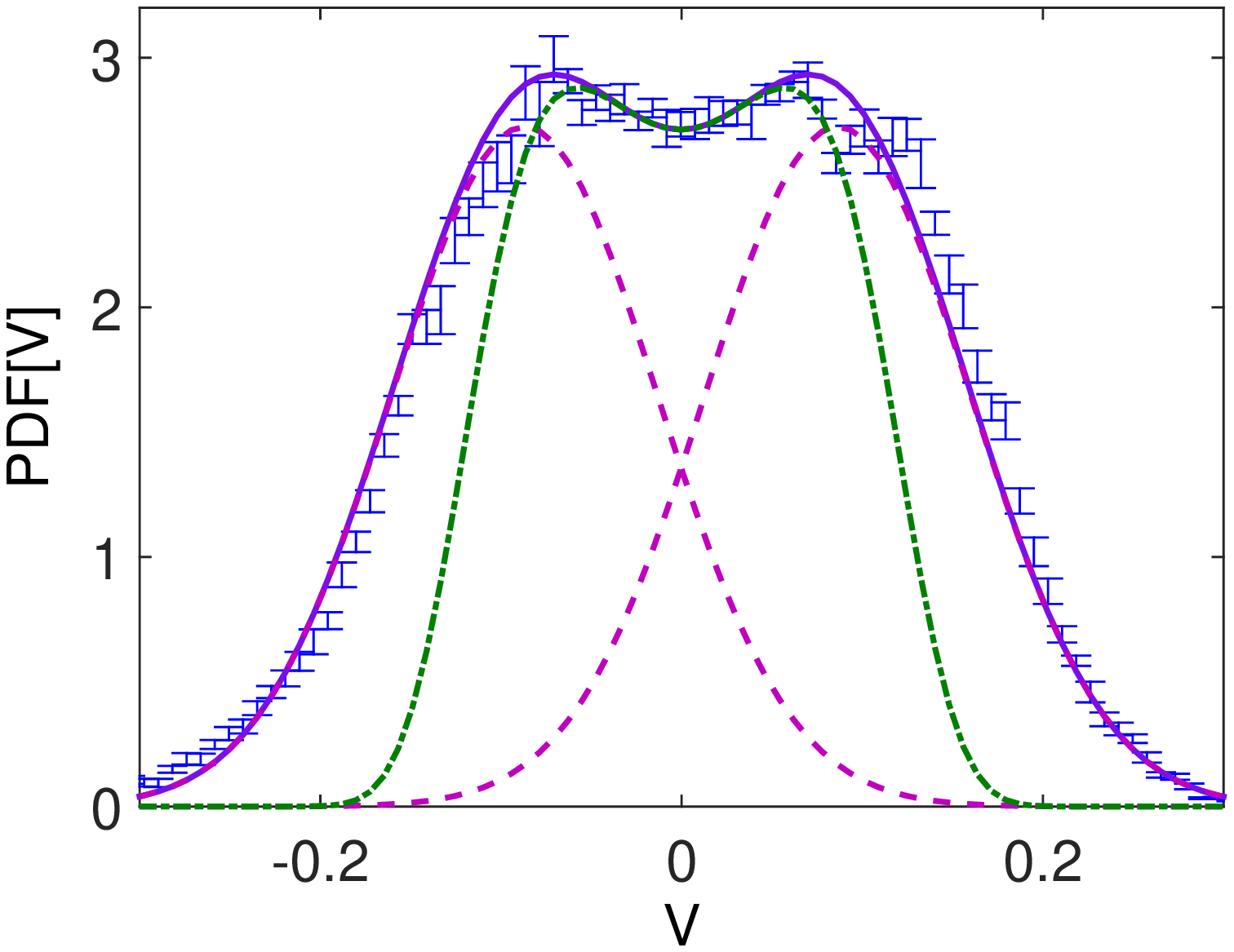}  
\end{center}
\caption{\label{figpdf}PDF of the amplitude of the large scale velocity for (from top to bottom) $Rh=12,14,20$. Symbols are experimental datas. The blue continuous curve is Eq. (\ref{defP1}), the sum of the two gaussians displayed with magenta dashed curves. The green dashed-dotted curve is Eq. \ref{eqlandau}.  For $Rh=12$, the two gaussians are very close to each other, so that the whole pdf is nearly gaussian. It becomes flatter close to $V=0$ at $Rh=14$ and it is bimodal at $Rh=20$. }
\end{figure}

\begin{figure}
\begin{center}
\includegraphics[width=75mm,height=60mm]{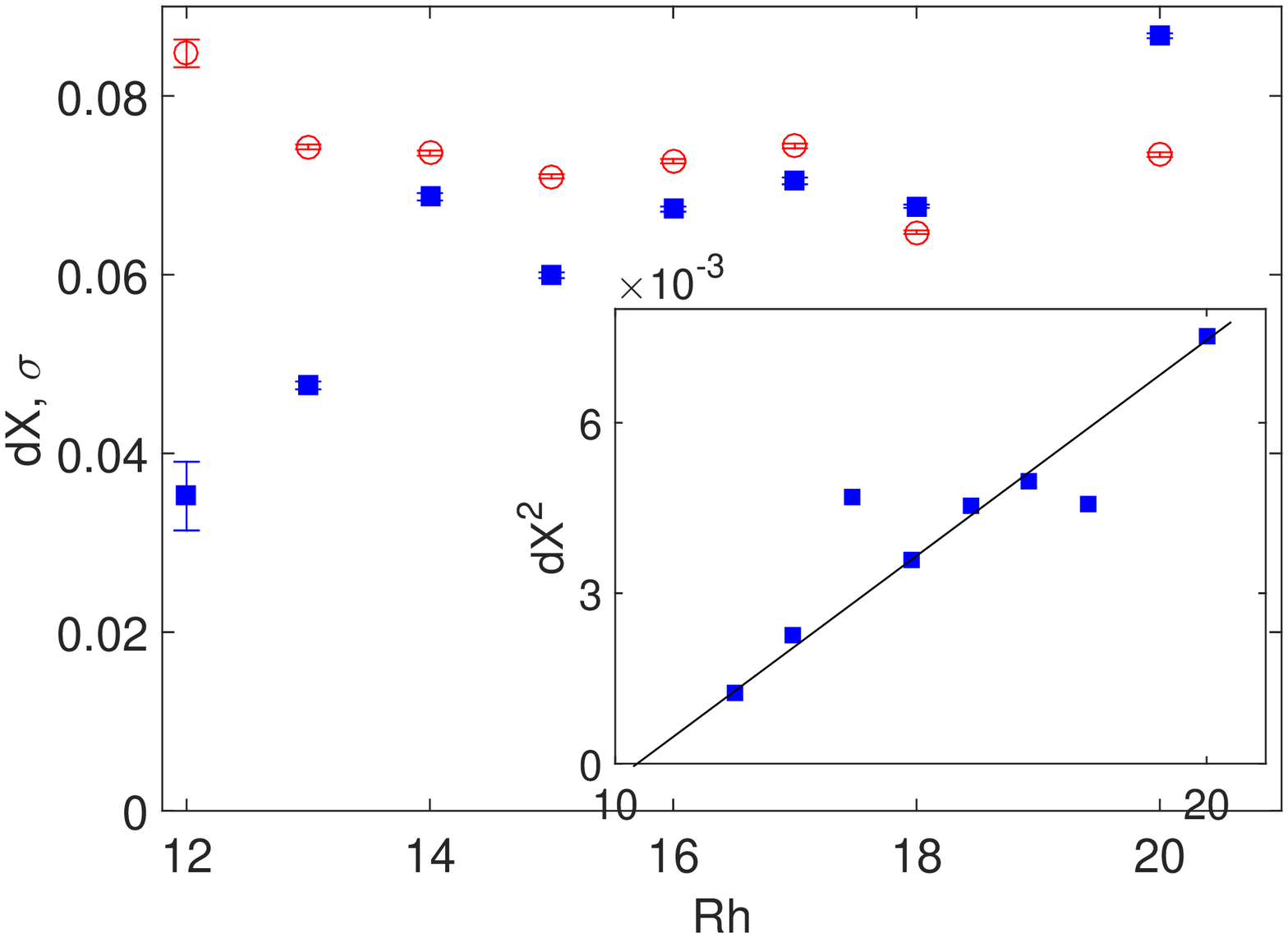}
\caption{\label{figumsigma}
 Parameter $dX$ ($\square$) and $\sigma$ ($\circ$) of Eq. (\ref{defP1}) as a function of $Rh$. Insert: same data, $dX^2$ as a function of $Rh$.}
\includegraphics[width=75mm]{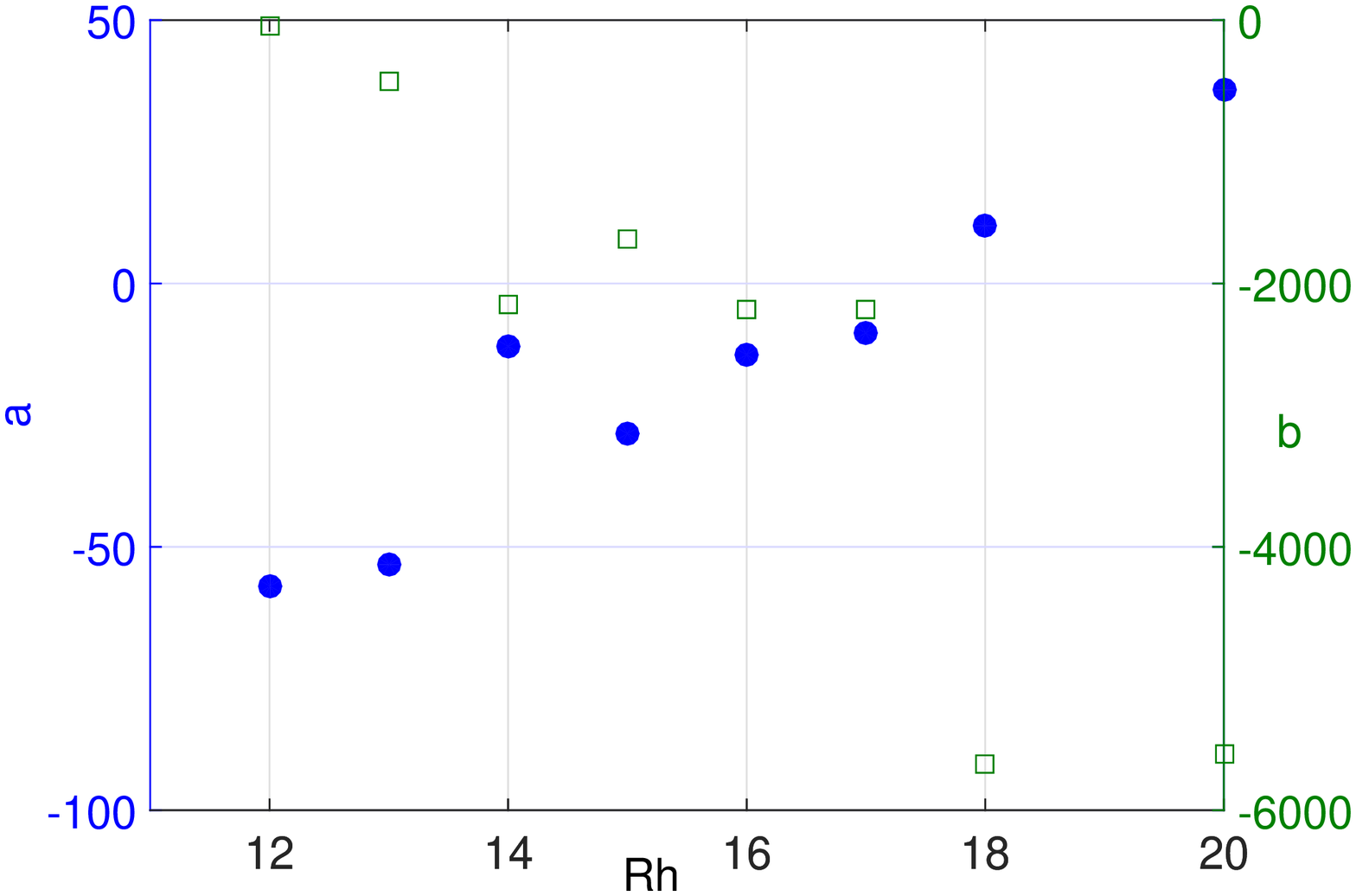} 
\caption{\label{figlandau} Parameter $a$ ($\circ$) and $b$ ($\square$) of Eq. (\ref{eqlandau}) as a function of $Rh$.}
\end{center}
\end{figure}

Extracting the values of $a$ and $b$ directly from the PDF results in
large error bars and numerical values that strongly depend  on the range 
over which the fit is achieved and on the possible asymmetry of the PDF.
This is not the case with the following data treatment : Eq. (\ref{defP1}) and (\ref{eqlandau}) are
 expanded close to $V=0$ and we then express $a$ and $b$ as a function of $\sigma$
and $dX$. The obtained values of $a$ and $b$ are displayed in fig. \ref{figlandau}. 
$a$ increases linearly with $Rh$ and changes sign in the vicinity of $Rh_7$.
This corresponds to the change of concavity of the PDF at $V=0$, and the
appearance of two non-zero maxima.
$b$ becomes very small below $Rh_6$, in agreement with the Gaussian behavior of
the PDF. It becomes more negative when $Rh$ increases.
 As displayed in fig. \ref{figpdf}, Eq. (\ref{eqlandau}) is a good fit of the PDF only for small
values of $V$: it does not describe the tails of the PDF.

Therefore, although Eq.  (\ref{eqlandau}) seems a more natural description for the transition of the PDF, it provides a less accurate fit than Eq. (\ref{defP1}) that is also valid for the tails of the velocity distribution.
Eq.  (\ref{eqlandau}) would be obtained for a system described by a free-energy proportional to $-a V^2- b V^4$ and subject to  additive fluctuations. In contrast Eq. (\ref{defP1}) is expected if $V$ is the sum of a constant velocity $\pm dX$ and random fluctuations of constant energy.   


When $Rh$ is further increased, the value of the PDF close to $V=0$ decreases. This corresponds to the
large scale circulation becoming more and more stable, and the reversals
between these two directions of the flow becoming less and less frequent.
Ultimately, no reversals are observed on the maximum measurement time 
(set by the stability of the experiment). The obtained PDF becomes
asymmetric and is peaked close to the  sign of rotation that is selected
initially.\\

{\bf Discussion}\\

We have studied the different bifurcations that the large scale flow undergoes. At small $Rh$, a sequence of bifurcation drives the system from zero large scale circulation to a steady non-zero one, then to a time-periodic state followed by a quasi-periodic one. When increasing $Rh$ further, the system becomes chaotic and explores successively positive and negative values of the large scale circulation. 

Other bifurcations then occur over a fluctuating background. Bifurcations in that context are by far less documented than bifurcations occurring over a steady or time-periodic state. Relevant quantities are statistical ones, such as moments or the PDF of the variable.

The PDF undergoes three bifurcations as $Rh$ increases. First, it becomes non Gaussian, with a Kurtosis departing from $3$. 
The whole PDF is then well described by the sum of two non-centered Gaussians. The distance between the center of the Gaussian increases while their standard deviation remains roughly constant, so that a second bifurcation occurs: the PDF becomes bimodal. We can describe this phenomenon using an analogy with Landau's theory: the behaviors of the parameters $a$ and $b$ follow Landau's assumption, which was not obvious since   this theory fails in low dimension when spatial fluctuations are of importance. Here the unstable mode is at large scale so that spatial coupling is not relevant. 

This evolution is associated in the time series to the appearance of long-lived coherent states during which the circulation does not change sign. The time series can then be described as random reversals between these two coherent states.
We had studied in detail the spectral properties of these time series \cite{jojoepl2015}. For $Rh$ between $10$ and $30$, the time series display $1/F$ noise. This behavior results from the distributions of the duration between sign changes that are heavily tailed. We note that $1/F$ fluctuations occur for $10<Rh<30$. They are therefore  observed for Gaussian, non Gaussian and bimodal PDF.

At even larger $Rh$, the mean duration between sign changes diverges. When it is larger than the duration of stability of the experiment (several hours), the observed PDF, measured over this maximum duration, will be restricted to positive or negative values (depending on the initial condition). 
We have thus observed in this experiment, two different scenarii that describe the disappearance of a regime of reversals.
At large $Rh$, reversals become less and less likely and eventually are no longer observed. The system remains stuck in one of the two states.
At low $Rh$, reversals disappear because the time series are so fluctuating, that one cannot identify anymore the two states connected by reversals. We expect that these two possible ways to destroy or create reversals are generic and are observed in other contexts \cite{MHD,RB}.

\end{document}